\documentstyle[aps,twocolumn,epsf]{revtex}
\begin{document}
\twocolumn[\hsize\textwidth\columnwidth\hsize\csname
@twocolumnfalse\endcsname
\title{Atomic Bose-Einstein Condensation with Three-Body
Interactions and Collective Excitations}
\author{A. Gammal$^{(1)}$, T. Frederico$^{(2)}$, Lauro Tomio$^{(1)}$
and Ph. Chomaz$^{(3)}$}
\address{$^{(1)}$ Instituto de F\'{\i}sica Te\'{o}rica,
Universidade Estadual
Paulista, \\
01405-900 S\~{a}o Paulo, Brazil \\
$^{(2)}$Departamento de F\'{\i }sica, Instituto Tecnol\'{o}gico da
Aeron\'{a}utica, \\
Centro T\'{e}cnico Aeroespacial, 12228-900 S\~{a}o Jos\'{e} dos Campos, SP,
Brazil \\
$^{(3)}$ GANIL, B.P. 5027, F-14021 Caen Cedex, France }
\date{\today}
\maketitle

\begin{abstract}
The stability of a Bose-Einstein condensed state of trapped ultra-cold atoms   
is investigated under the assumption of an attractive two-body and
a repulsive three-body interaction.
The Ginzburg-Pitaevskii-Gross (GPG) nonlinear Schr\"odinger equation is
extended  to include an effective potential dependent on the square of the
density and solved numerically for the $s-$wave.
The lowest frequency of the collective mode is determined 
and its dependences on the number of atoms and on
the strength of the three-body force are studied.
We show that the addition of three-body dynamics can allow the number
of condensed atoms to increase considerably,
even when the strength of the three-body force is very small
compared with the strength of the two-body force. We also observe a
first-order liquid-gas phase transition for the condensed
state up to a critical strength of the effective three-body force.
\end{abstract}
\vskip 0.5cm ]

\section{Introduction}

The theoretical research on Bose-Einstein condensation (BEC)~\cite{bec}, a
phenomenon predicted more than 70 years ago, is receiving considerable
experimental and theoretical support in recent years~\cite{PW}.
The relevance of BEC for understanding the properties of liquid $^4$He
was pointed out by London~\cite{london}, suggesting that the 
peculiar phase transition that liquid helium undergoes at 2.18K is 
a BEC phenomenon. It is also important to observe that, at the level of
two-body collisions, Bogoliubov in 1947~\cite{bogo} has shown for homogeneous
gas that BEC is only possible for systems with repulsive potentials.

Intense experimental researches on BEC for magnetically
trapped weakly interacting atoms have been done 
recently~\cite{and95,brad97,mew95,hidr}. 
In the experiment reported in \cite{and95}, a condensate of approximately
2000 spin-polarized $^{87}$Rb atoms was produced in a cylindrically 
symmetric magnetic trap~\cite{PW,cornell}. \ 
It is a common understanding that, in low temperature and density,
where interatomic distances are much greater than the distance scale of
atom-atom interactions, two-body interactions take a simple form, and
three-body interactions can be neglected. \ At such regime, only two-body
$s-$wave scattering is important.  With temperature low enough 
the magnitude of
the scattering length $a$ is much less than the corresponding thermal 
de Broglie wavelength, and the exact shape of two-atom interaction is
unimportant.

The experimental evidences of Bose-Einstein condensation (BEC) in
magnetically trapped weakly interacting atoms~\cite{and95,brad97,mew95,hidr}
brought a considerable support to the theoretical research on bosonic
condensation. The nature of the effective atom-atom interaction determines the
stability of the condensed state: the two-body pseudopotential is repulsive for
a positive $s-$wave atom-atom scattering length and it is attractive for a 
negative scattering length~\cite{huang}. The ultra-cold trapped atoms
with repulsive two-body interaction undergoes a phase-transition to a
stable Bose condensed state, in several cases found experimentally, as
for $^{87}$Rb~\cite{and95}, $^{23}$Na~\cite{mew95} and $^1$H~\cite{hidr}. 
However, a condensed state of atoms with negative $s-$wave atom-atom scattering
length (as in case of $^{7}$Li~\cite{brad97}) would be unstable, unless the
number of atoms $N$ is small enough such that the stabilizing force provided by the
harmonic confinement in the trap overcomes the attractive interaction, as
found on theoretical grounds~\cite{rup95,baym96}. It was indeed observed
in the $^{7}$Li gas~\cite{brad97}, for which the $s-$wave scattering length 
is $a=-14.5\pm 0.4$ \AA , that the number of allowed atoms in the Bose 
condensed state was limited to a maximum value between 650 and 1300, which
is consistent with the mean-field prediction~\cite{rup95}. 

So, for systems of atoms with attractive two-body interaction, 
it is widely believed~\cite{rup95,bag99,kagan96} that the condensate has no 
stable solution above certain critical number of atoms $N_{max}$. 
However, in this case the addition of a repulsive potential derived from 
three-body interaction is consistent with a number of atoms larger than $N_{max}$. 
Even for a very small strength of the  three-body force, the region of stability for the
condensate can be extended  considerably, as previously reported in \cite{ADV,GFT}, and
shown in more detail in the present work.  By considering the possible effective
interactions, it was reported in Ref.~\cite{esry} that a sufficiently
dilute and cold Bose gas exhibits similar three-body dynamics for both
signs of the $s-$wave atom-atom scattering length. 
It was also suggested that, for a large number of bosons the three-body
repulsion can overcome the two-body attraction, and a stable condensate
will appear in the trap~\cite{josse}.
If an atomic system is characterized by having effectively an attractive 
two-body interaction together with a repulsive three-body interaction,
two mechanisms for stability are possible: 
(a) the kinetic energy acting at lower densities and 
(b) the repulsive weak three-body force effective at higher densities.
These mechanisms indicate that, for the same number of atoms, one
lower-density phase and a higher-density phase can be found, if the 
three-body force is weak enough not to dominate the effective 
interaction. 

It was pointed out in Ref.~\cite{kagan} that an easier experimental
approach to
probe density fluctuations is to consider an observable directly sensitive to
the probability of finding three atoms near each other, which will correspond
to the loss rate of atoms due to three-body recombination. Such a three-body
recombination rate in BEC, was considered recently in Refs.~\cite{fedi},
\cite{burt} and \cite{kagan96} (see also the review of \cite{bag99}).
It was shown in Ref.~\cite{fedi} that the three-body recombination
coefficient of ultracold atoms to a weakly bound $s$ level goes to infinity
in the Efimov limit~\cite{efimov}. The Efimov limit is a particularly
interesting three-body effect, which happens when the two-body scattering
length is very large (positive or negative). In this case, with the two
boson energy close to zero, the three-boson system presents an increasing
number of loosely bound three body states, which have large spatial
extension and do not depend on the details of the interaction~\cite{3atom}. 
So, our main motivation here is to provide an extension to the GPG 
equation~\cite{gin,Pita}, which considers a three-body interaction and, 
in this way, provides the framework for a numerical investigation of the
relevance of three-body interaction in Bose-Einstein condensation. 

In the present work we consider a possible general scenario of atomic systems
with attractive two-body and repulsive three-body interactions. By using the 
mean-field approximation, we investigate the competition between the leading
term of an attractive two-body interaction, originated from a negative two-atom
$s-$wave scattering length, and a repulsive three-body effective interaction,
which can happen in the Efimov  limit~\cite{efimov}
($|a|\rightarrow \infty$) as discussed in  Ref.~\cite{esry}~\footnote{The
physics of three-atoms in the Efimov limit is
discussed in Ref.~\cite{3atom}, that extends a previous study of universal
aspects of the Efimov effect~\cite{halo}.}. 
We show that, in a dilute gas, a small repulsive three-body force
added to an attractive two-body interaction is able to stabilize the
condensate beyond the critical number of atoms in the trap, found just
with attractive two-body force~\cite{rup95}, such that a kind of
liquid-gas phase-transition occurs. The plan of the paper is as follows.
In section II, we introduce the Ginzburg - Pitaevskii - Gross (GPG)
formalism. In section III, we present the main numerical results for the
static solutions, together with a variational analysis. In section IV, we
present a stability analysis and results for collective excitation in the
condensate. In this section IV we also observe that the inclusion of three 
body effects points out possible evidences of a liquid - gas phase 
transition in the condensate.  Finally, in section V, we present our main
conclusions.

\section{Ginzburg - Pitaevskii - Gross Formalism}

In the following, we present our formalism, where the original
Ginzburg - Pitaevskii - Gross (GPG) non-linear equation~\cite{gin,Pita}, 
which includes a term proportional to the density (two-body interaction), is
extended through the addition of a term proportional to the squared-density
(three-body interaction). Next, after reducing such equation to
dimensionless units, we study  numerically the $s-$wave solution by varying
the corresponding dimensionless parameters, which are related to the
two-body scattering length, the strength of the three-body interaction and
the number of atoms in the condensed state. 
As particularly observed in Ref.~\cite{hs}, to incorporate all two-body
scattering processes in such many particle system, the two-body potential
should be replaced by the many-body $T-$matrix. \ \ Usually, at very low
energies, this is approximated by the two-body scattering matrix, which is
directly proportional to the scattering length~\cite{baym96}. So, in
order to obtain the desired equation, we first consider the effective
Lagrangian, which describes the condensed wave-function in the Hartree
approximation, implying the GPG energy functional: 

\begin{eqnarray}
{\cal {L}}&=&\int d^3r \left[ \frac{i\hbar}2\Psi^{\dagger}(\vec r) 
\frac{\partial \Psi(\vec r)}{\partial t}-\frac{i\hbar}2\frac{\partial
\Psi^{\dagger}(\vec r)}{\partial t}\Psi(\vec r)
+ \right. \nonumber \\ && \left. 
+ \frac{ \hbar ^2}{2m} \Psi^{\dagger}(\vec r)\nabla ^2\Psi(\vec r) 
- \frac m2 \omega^2 r^2 |\Psi(\vec r)|^2\right] +{\cal {L}} _{{\rm 
{I}}}\ .  \label{lag}
\end{eqnarray}
In our description, the atomic trap is given by a rotationally symmetric
harmonic potential, with angular frequency $\omega$, and ${\cal {L}}_{{\rm 
{I}}}$ gives the effective atom interactions up to three particles.

The effective interaction Lagrangian for ultra-low temperature bosonic
atoms, including two- and three-body scattering at zero energy, is written
as: 
\begin{eqnarray}
{\cal L}_{{\rm I}} &=& -\frac{1}{2}\int
d^3r_1d^3r_2d^3r^{\prime}_1d^3r^{\prime}_2 \Psi^\dagger 
(\vec{r^{\prime}}_1)\Psi^\dagger (\vec{r^{\prime}}_2) \Psi (\vec{r}_1) 
\Psi(\vec{r}_2) 
\nonumber \\ &&\times 
\left\langle \vec{r^{\prime}}_{12} \left| T^{(2)}(0) \right| 
\vec{r}_{12} \right\rangle \delta^3(\vec{r}_1^\prime+\vec{r}_2^\prime-
\vec{r}_1-\vec{r}_2)  
\nonumber \\ &&
-\frac{1}{3{\rm {!}}}\int \prod_{i=1}^3 \left( d^3r_i
d^3r^{\prime}_i
\Psi^{\dagger}(\vec{r}_i^{\prime}) 
\Psi (\vec{r}_i) \right)
\delta^3 ( \vec{R}_{123}^{\prime}-\vec{R}_{123} ) 
\nonumber \\ && \times 
\left\langle\vec{r}_{12}^{\prime}\vec{R}_3^{\prime}
\left|T^{(3)}(0)- \sum_{j<k}T^{(2)}_{jk} (0- K_i)\right|\vec{r}_{12}
\vec{R}_3\right\rangle 
\ ,  \label{li}
\end{eqnarray}
where $\vec r_{12}$ and $\vec R_3$ are the relative coordinates, given by 
$\vec r_{12}=\vec r_1-\vec r_2$ and $\vec R_3= \vec r_3-(\vec r_1+\vec r_2)/2$;
and $\vec R_{123}\equiv (\vec r_1+\vec r_2+\vec r_3)$. \ \ $T^{(3)}(0)$
and $T^{(2)}_{jk}(0)$ are the corresponding three-body $T-$matrix and
two-body $T-$matrix for the pair ${jk}$, which are evaluated at zero-energy.
The two-body $T-$matrix for each pair $(jk)$ is subtracted from $T^{(3)}(0)$
to avoid double counting and $K_i$ is the kinetic energy operator for
particle $i$. \ 

We can approximate the above effective interaction Lagrangian at low
densities by averaging the $T-$matrices over the relative coordinates,
considering that the thermal wave-length is much greater than the
characteristic interaction distances. 
\begin{eqnarray}
&&{\cal {L}}_{{\rm I}}=-\frac{1}{2}\int d^3r^{\prime}_{12}d^3r_{12}
\left\langle \vec{r^{\prime}}_{12} \left| T^{(2)}(0) \right| \vec{r}_{12}
\right\rangle \int d^3r \left|\Psi (\vec{r})\right|^4  \nonumber \\
&-&\frac{1}{3 {\rm {!}}}\int d^3r^{\prime}_{12} d^3R^{\prime}_3d^3r_{12}
d^3R_3 
\int d^3r \left|\Psi (\vec{r})\right|^6 
\nonumber \\ &\times & 
\left\langle \vec{r^{\prime}}_{12}\vec{R^{\prime}}_3 \left|
T^{(3)}(0)-\sum_{j<k}T^{(2)}_{jk}(0- K_i) \right| \vec{r}_{12}\vec{R}_3
\right\rangle  \ .  \label{li1} \end{eqnarray}
The integrations of the $T$-matrices over the relative coordinates gives the
zero momentum matrix elements: 
\begin{eqnarray}
&&\int d^3r^{\prime}_{12}d^3r_{12} \left\langle \vec{r^{\prime}}_{12}\left|
T^{(2)}(0)\right|\vec{r}_{12}\right\rangle = 
\nonumber \\ &&
(2 \pi)^3 \left\langle 
\vec p_{12}=0 \left| T^{(2)}(0)\right|\vec{p}_{12}=0 \right\rangle  
= \frac{4\pi\hbar^2 a}{m} \ ,  \label{t2}
\end{eqnarray}
where $a$ is the two-body scattering length. For the connected three-body 
$T-$matrix, also by integrating over the coordinates, we obtain the 
 corresponding zero momentum ($\vec{p}_{12}=0, \vec{P}_3=0$) matrix
elements, which give us the strength of the three-body effective
interaction $\lambda_3$, as follows: 
\begin{eqnarray}
&&\int d^3r^{\prime}_{12} d^3R^{\prime}_3d^3r_{12} d^3R_3 
\nonumber \\ &&\times
\left\langle \vec{r^{\prime}}_{12}\vec{R^{\prime}}_3 
\left|T^{(3)}(0)-\sum_{j<k}T^{(2)}_{jk}(0- K_i) \right| 
\vec{r}_{12}\vec{R}_3 \right\rangle  = \nonumber \\
&&= (2 \pi)^6 \left\langle \left|
T^{(3)}(0)-\sum_{j<k}T^{(2)}_{jk}(0- K_i) \right|\right\rangle  
\nonumber \\&&
= 2 \lambda_3 \ ,  \label{t3}
\end{eqnarray}
where $\langle |\equiv \langle{\vec{p}_{12}=0,\vec{P}_3=0}|$
and $|\rangle \equiv |{\vec{p}_{12}=0,\vec{P}_3=0}\rangle$.

The nonlinear Schr\"{o}dinger equation, which describes the condensed
wave-function in the mean-field approximation, is obtained from the
effective Lagrangian given in Eq.~(\ref{lag}). By considering the
interaction in Eq.~(\ref{li1}), it can be written as~\cite{fw} 
\begin{eqnarray}
i\hbar \frac{\partial \Psi (\vec{r},t)}{\partial t} &=&
\left[-\frac{\hbar ^{2}}{2m}\nabla ^{2}+\frac{m}{2}\omega ^{2}r^{2}
-N\frac{4\pi \hbar ^{2}|a|}{m}|\Psi (\vec{r},t)|^{2}
 \right. \nonumber \\ &+& \left.
\lambda _{3}N^{2}|\Psi (\vec{r},t)|^{4}\right] 
\Psi (\vec{r},t).  \label{NLS}
\end{eqnarray}
For a stationary solution, $\Psi (\vec{r},t)=e^{-i\mu t/\hbar }$ $\psi 
(\vec{r})$, and the above equation can be written as 
\begin{eqnarray}
\mu \psi (\vec{r})&=&\left[ -\frac{\hbar ^{2}}{2m}\nabla ^{2}+\frac{m}{2}
\omega ^{2}r^{2}-N\frac{4\pi \hbar ^{2}|a|}{m}|\psi (\vec{r})|^{2}
\right.\nonumber \\ &+&\left.
\lambda_{3}N^{2}|\psi (\vec{r})|^{4}\right] \psi (\vec{r}),  \label{SNLS}
\end{eqnarray}
where $\mu $ is the chemical potential (single particle energy) and $\psi 
(\vec{r})$ is normalized as 
\begin{equation}
\int d^{3}r|\psi (\vec{r})|^{2}\ =\ 1.  \label{norm1}
\end{equation}
The total energy of the system is given by 
\begin{eqnarray}
E&=&\int d^{3}r\left\{ N\frac{\hbar ^{2}}{2m}\left| \nabla \psi (\vec{r}
)\right| ^{2}+N\frac{m}{2}\omega ^{2}r^{2}\left| \psi (\vec{r})\right| ^{2}
\right. \nonumber \\ &-& \left.
\frac{N^{2}}{2}\frac{4\pi \hbar ^{2}|a|}{m}|\psi (\vec{r})|^{4}+\frac{N^{3}}
{3}\lambda _{3}|\psi (\vec{r})|^{6}\right\} .  \label{Etot}
\end{eqnarray}
The central density of the system can be obtained directly from the solution
of the above equation, normalized as in Eq.~(\ref{norm1}): 
\begin{equation}
\rho _{c}=N|\psi (0)|^{2}.  \label{rho}
\end{equation}
The physical scales presented in the above equations can be easily
recognized by working with dimensionless equations. By rescaling
Eq.~(\ref{SNLS}) for the $s-$wave solution, we obtain 
\begin{equation}
\left[ -\frac{d^{2}}{dx^{2}}+\frac{1}{4}x^{2}-\frac{|\Phi (x)|^{2}}{x^{2}}
+g_{3}\frac{|\Phi (x)|^{4}}{x^{4}}\right] \Phi (x)\ =\ \beta \Phi (x)\ ,
\label{schd}
\end{equation}
where $x\equiv \sqrt{{2m\omega }/{\hbar }}\ r$ and $\Phi (x)\equiv N^{1/2} 
\sqrt{8\pi |a|}\;r\psi (\vec{r})$. The dimensionless parameters, related to
the chemical potential and the three-body strength are, respectively, given
by 
\begin{equation}
\beta \equiv \frac{\mu }{\hbar \omega }\;\;\;{\rm and}\;\;\;g_{3}\equiv
\lambda _{3}\hbar \omega \left[ \frac{m}{4\pi \hbar ^{2}a}\right] ^{2} .
\label{beta-g}
\end{equation}
The normalization for $\Phi (x)$, obtained from Eq.~(\ref{norm1}), defines
a number $n$ related to the number of atoms $N$: 
\begin{equation}
\int_{0}^{\infty }dx|\Phi (x)|^{2}\ =n,\;\;\;{\rm {where}}\;\;\;n\equiv
2N|a| \sqrt{\frac{2m\omega }{\hbar }}\ .  \label{norm2}
\end{equation}
The boundary conditions in Eq.(\ref{schd}) are given by~\cite{rup95} 
\begin{eqnarray}
\Phi(x\to 0) &\to& 0 \nonumber \\ 
\Phi(x\to \infty) &\propto&  
\exp\left(-\frac{x^2}{4}+\left[\beta-\frac{1}{2}\right]\ln (x)\right)
.\label{boundary} \end{eqnarray}
In terms of the dimensionless variables, the total energy of the system is
given by 
\begin{eqnarray}
E&=& {\hbar \omega N} \int_0^\infty dx\left\{ \left| \frac{d\phi(x)}{dx}
\right|^{2}+\frac{x^2\phi^2(x)}{4}
\right.\nonumber\\ &-&\left. 
-\frac{n\phi^4(x)}{2 x^{2}}+\frac{
n^2g_3\phi^6(x)}{3x^4}\right\} ,  \label{Etot2}
\end{eqnarray}
where $\phi (x)\equiv\Phi (x)/n^{1/2}$ is normalized to one.

\section{Liquid-Gas Phase Transition - Static Solutions}

\subsection{Variational Approach} 
As a further reference to our results, and the stability analysis,  
it will be helpful first to consider a variational procedure~\cite{fetter}, 
using a trial gaussian wave function for $\psi (\vec{r})$.
So, in Eq.~(\ref{Etot}) we consider the following
variational wave function (normalized to one): 
\begin{equation}
\psi _{var}(\vec{r})=\left( \frac{1}{\pi \alpha ^{2}}\frac{m\omega }{\hbar }
\right) ^{\frac{3}{4}}\exp {\left[ -\frac{r^{2}}{2\alpha ^{2}}\left( \frac{
m\omega }{\hbar }\right) \right] },  \label{varwf}
\end{equation}
where $\alpha $ is a dimensionless variational parameter. The corresponding
root-mean-\-square radius, $r_{0}$, will be proportional to the variational
parameter $\alpha$: 
\begin{equation}
r_{0}\equiv \sqrt{\langle r^{2}\rangle _{var}}=\alpha \sqrt{\frac{3\hbar}
{2m\omega }}\;.  \label{msqr}
\end{equation}
The expression for the total variational energy, which is obtained after
replacing Eq.~(\ref{varwf}) in Eq.~(\ref{Etot}), is given by 
\begin{equation}
E_{var}(\alpha )=\hbar \omega N\left[ \frac{3}{4}\left( \alpha ^{2}+\frac{1}{
\alpha ^{2}}\right) -\frac{n}{4\sqrt{\pi }\alpha ^{3}}+\frac{2n^{2}g_{3}}{9
\sqrt{3}\pi \alpha ^{6}}\right] .  \label{Evar}
\end{equation}
In the same way, we can obtain the corresponding variational expression for
the single particle energy, Eq.~(\ref{SNLS}): 
\begin{equation}
\mu _{var}(\alpha )=\hbar \omega \left[ \frac{3}{4}\left( \alpha ^{2}+\frac{1
}{\alpha ^{2}}\right) -\frac{n}{2\sqrt{\pi }\alpha ^{3}}+\frac{2n^{2}g_{3}}{3
\sqrt{3}\pi \alpha ^{6}}\right] .  \label{muvar}
\end{equation}
The variational central density, using Eqs.~(\ref{rho}) and (\ref{varwf}),
can also be given in terms of this parameter $\alpha $: 
\begin{equation}
\rho _{c,var}(\alpha )=\left( \frac{m\omega }{\pi \hbar }\right) ^{3/2}\frac{
1}{\alpha ^{3}}.  \label{rhovar}
\end{equation}
The approximate solutions for the total energy are obtained from the 
extrema of (\ref{Evar}) with respect to variation of the parameter $\alpha$.

The variational solutions of $E_{var}(\alpha )$ are given, as a function
of $n$ and $g_{3}$ (where $a<0$ and $g_{3}>0$), by the real roots of
$\partial E_{var}(\alpha )/\partial\alpha =0$~\footnote{By using a 
numerical procedure one can reach easily the extrema of Eq.~(\ref{Evar})
by varying the parameter $\alpha$, once the other parameters are fixed.}.

\begin{figure}
\setlength{\epsfxsize}{1.0\hsize}\centerline{\epsfbox{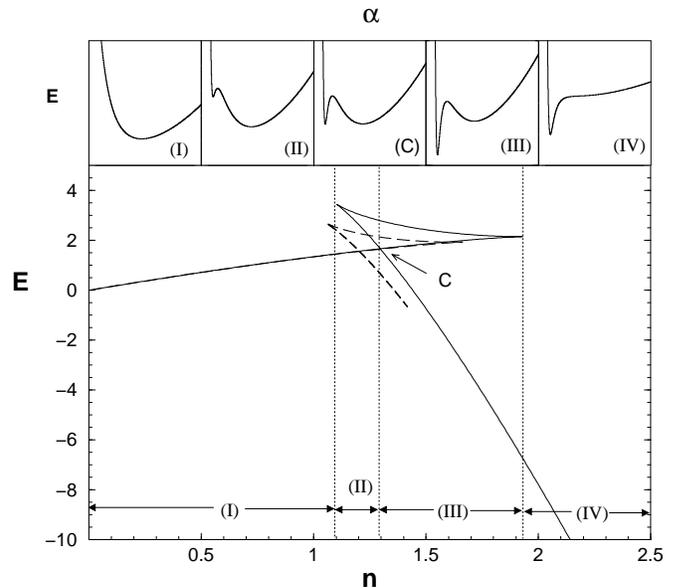}}
\caption[dummy0]{
In the lower part, we have a comparison between variational (solid curve)
and exact (dashed curve) numerical calculations of the condensate energy
as a function of the reduced number of atoms $n$ for $g_{3}=0.005\ .$ In the
upper frame we show five  plots of the variational energy as a function
of the variational parameter $\alpha $ for  five particular values of $n$
shown also in the lower frame. (I) (resp IV) corresponds to a small (large)
$n$ region where only one stable solution is encountered; (II) (resp III) to
a small (large) $n$ region where we observe three extrema for the energy;
(C) corresponds to a particular $n$ where we obtain two stable solutions
with the same energy $E_{1}=E_{2}$.  $E$ is given in units of $(N\hbar
\omega )/n$.}
\end{figure}
In Fig. 1, we first illustrate the variational procedure
considering an arbitrarily small three-body interaction, chosen as $
g_{3}=0.005$. In the upper part of the figure, we show five small plots
for the total variational energy $E$, in terms of the variational width
$\alpha $. Each one of the small plots corresponds to particular values
of $n$. For each number $n$ we  report the energy of the variational
extrema in the lower part of Fig. 1. In region (I) where the number of
atoms is still small, the attractive two body force dominates over the
repulsive three-body force and just one minima of the energy as a
function of the variational parameter $\alpha $ is found.
That is also the case for $g_{3}=0$.
When the number of atoms is further increased (region (II)) two minima
appear in the energy $E\left( \alpha \right) .$ An unstable maximum
is also found between the two minima. The lower energy minimum is stable
while the solution corresponding to the smaller $\alpha $ is metastable.
This solution has a higher density and, consequently, its metastability
is justified by the repulsive three-body force acting at higher
densities. The minimum number $n$ for the appearance of the metastable
state is characterized by an inflection point in the energy as a function
of $\alpha $. The value of $n$ at the inflection
point corresponds to the beak in the plot of extremum energy versus $n$
because for larger $n$ three variational solutions are found as depicted
in the lower part of Fig. 1. The attractive two-body and trap
potentials dominate the condensed state in the low-density stable phase
up to the crossing point (C). At this point, the denser metastable
solution becomes degenerate in energy with the lower-density stable
solution and a first order phase transition takes place. Since the two
solutions differ by their density this transition is analogous to a
gas-liquid phase transition for which the density difference between the
liquid and the gas is the order parameter. In the variational calculation
this occurs at the transition number $n\approx$1.3 while the numerical
solution of the NLSE gives $1.2$. In region (III), we observe two local
minima with different energies, a higher-density stable point and a
lower-density metastable point. The metastable solution disappears in the
beak at the boundary between region (III) and (IV). In regions (III) and
(IV) the three-body repulsion stabilized a dense solution against the
collapse induced by the two-body attraction. The qualitative features of
the variational solution is clearly verified by the numerical solution of
the NLSE, as shown by the dashed curve.

\subsection{Numerical Results}

The numerical solutions of Eq.~(\ref{schd}) are obtained for several values
of $\beta$, using three values of $g_3$ to characterize the solutions. We
have used the Runge-Kutta (RK) and ``shooting" method to obtain the
corresponding solutions in each case~\cite{num}. 
The stability assignment for the stationary solutions was made by
studying the corresponding time dependent Schr\"{o}dinger equation, using the
Crank-Nicolson (CN) method (see Refs.~\cite{rup95} and \cite{ames}). The
numerical procedure to determine such stability was done in the following way:
when applying the CN method, we started by using the static solution obtained
from the RK method and observed if the modulus of the wave function remained
constant. If this was occurring for a long period of time (of about 500 units
of dimensionless time $\tau = \omega t$) the solution was considered stable,
otherwise unstable. 

In Fig. 2 we present the total energy as a function of the number of 
atoms, represented by the reduced number $n$ defined in Eq.~(\ref{norm2}),
for three significative values of the quintic parameter $g_{3}$,
given by 0, 0.016 and 0.03. The results agree with Ref.\cite{ADV}.
When $g_3=0$, the stable solutions for the energy starts at zero (for $n=0$)
and reaches a critical limit at $n_{max}\simeq 1.62$. There is no solutions for
higher $n$, but the plot also shows a branch with unstable solutions 
(with higher energies) for $n\le 1.62$. Our results are consistent with
results given in Ref.~\cite{hs}.
When $g_3=0.03$, only stable solutions appear for the energy, with no 
limit in the number of atoms, having a maximum at $n\sim 2$. So, this 
and higher values for $g_3$ already represent a dominance of the quintic term
in the interaction of Eq.~(\ref{schd}). We observe that the numerical 
stability analysis is consistent with the variational approach discussed
in the previous sub-section. 
The more interesting case  represented in Fig. 2 is for $g_3=$0.016, as in such
a case we observe a region of the plot where we can have up to three solutions
for the same $n$. The inset to this figure amplifies the region of the plot
where, for $g_3=0.016$, the solutions become unstable (between A and B) or 
metastable (between A and C, or B and C). At the point C a phase transition
occurs from a less denser (gas) to a more denser (liquid) phase.
\begin{figure}
\setlength{\epsfxsize}{1.0\hsize}\centerline{\epsfbox{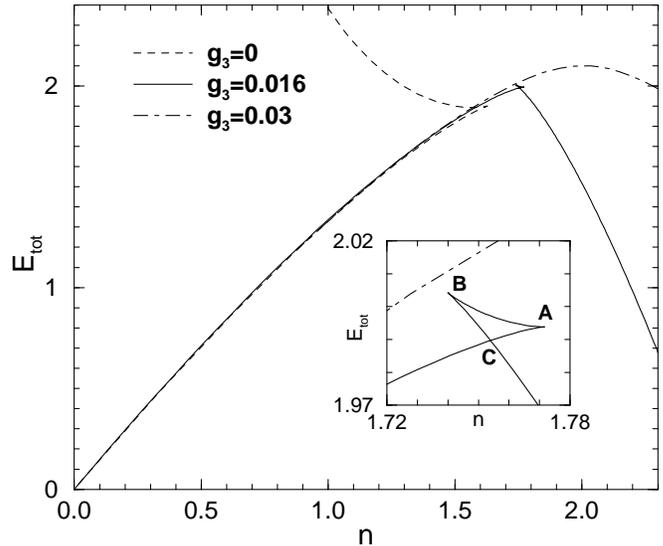}}
\caption[dummy0]{The total energy, in units of ${(\hbar\omega)}/\left(2|a|
\sqrt{(2m\omega)/\hbar}\right)$, is shown as a function of the reduced
number of atoms $n$, given by Eq.~(\ref{norm2}), for 
$g_3=$0, 0.016 and 0.03. 
The inset points out critical limits discussed in the text.}
\end{figure}

In Fig. 3, following a correspondence to  Fig. 2, we present the
results for the chemical potential in dimensionless units ($\beta$) as a 
function of $n$. The line with arrow in the inset to this figure 
indicates the approximate position in $n$, where the phase-transition 
(from a `gas' phase to a `liquid' phase) occurs.
For $g_{3}=0.016$ the part of the
plot linking points A and B is unstable (see both Figs. 2 and 3),
otherwise it is stable. Finally, for $g_{3}=0.03$, the function of the
energy in terms of $n$ is always single valued and stable.
Our calculation for $g_{3}=0$ also agrees with results presented in
Ref.~\cite{rup95}, with the maximum number of atoms limited to
$n_{max}\approx 1.62$~\footnote{Our $n$ is equal to $|C_{nl}^{3D}|$ of 
Ref.~\cite{rup95}.}.  As we can see, for $n\leq n_{max}$ two solutions are
possible, one of them being unstable. \
For $g_{3}$ higher than zero, a new pattern appears.
For instance, the plot for the case of $g_{3}=0.016$ (see the inset) can be 
divided in several sectors according to the stability
analysis, with the help of Fig. 2: Starting from $n=0$ ($\beta =1.5$)
until point C$_{\rm G}$, and from C$_{\rm L}$ to higher values of $n$, we have stable
solutions; from C$_{\rm G}$ to A and from B to C$_{\rm L}$ we obtain metastable solutions;
from A to B the solutions are unstable, corresponding to maxima for the 
energies.

\begin{figure}
\setlength{\epsfxsize}{1.0\hsize}
\centerline{\epsfbox{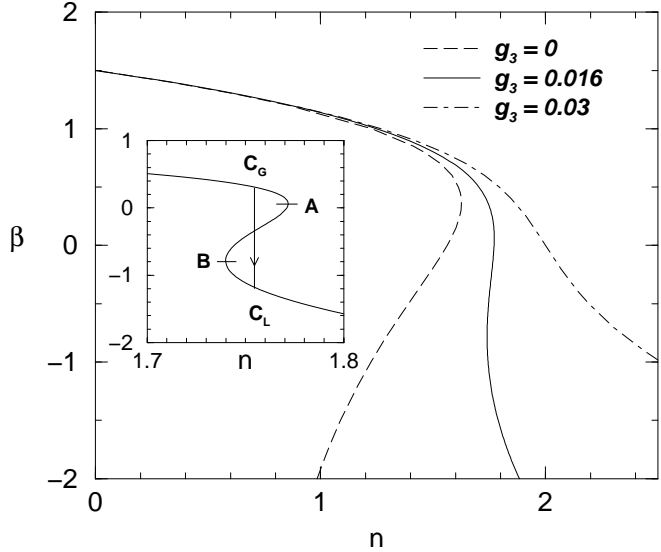}} 
\caption[dummy0] {The chemical potential, in dimensionless units
($\beta=\mu/(\hbar\omega)$, is shown as a function of the reduced 
number of atoms $n$, for the same set of $g_3$ shown in Fig. 2. 
The inset points out the critical limits corresponding to Fig. 2
(C$_{\rm G}$ and C$_{\rm L}$ corresponds to C), and the straight line with arrow
indicates the transition from a less denser to a more denser phase.}
\end{figure}

In Fig.4 we also plot the central density $\rho_c$, defined in
Eq.~(\ref{rho}),
as a function of the number $n$. We use the same values of the
parameter $g_3$ as used in Figs. 2 and 3. The inset to the figure also 
points out the phase transition which occurs when $g_3=$0.016. 
As the straight line with arrow shows, after the transition the system
becomes more than three times denser than the original one.
Also, for $0<\ g_{3}\ <\ 0.0183$, we observe that the density $\rho _{c}$
presents back bending typical of a first order phase transition. 

By extending the observations of a first order phase transition, given in 
Figs. 2-4 for $g_3=$0.016, we also determined the region of $g_3$ where
such  kind of phase-transition can occur.
In Fig. 5 we have a phase-diagram, where it was shown the critical boundary
separating the two phases and a critical point at $n=1.8$ and $g_3=0.0183$. 
For $g_3$ less then such critical value, we observe 
two regions with distinct phases, similar to gas and
liquid phases.  These two different phases are also 
clearly identified in our Fig. 4, where we present the central
density as a function of $n$. 

\begin{figure}
\setlength{\epsfxsize}{1.0\hsize}
\centerline{\epsfbox{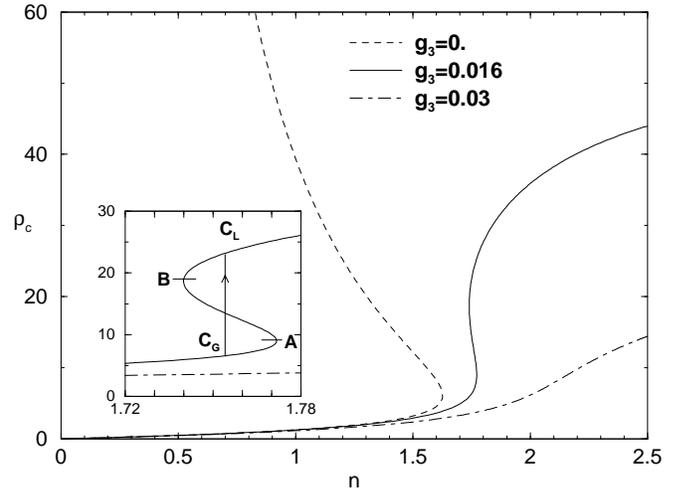}} 
\caption[dummy0] {Central density, in dimensionless units, as a function of
the number $n$, for the same set of parameters $g_3$ given in Figs. 2 and
3.}
\end{figure}
\begin{figure}
\setlength{\epsfxsize}{1.0\hsize}
\centerline{\epsfbox{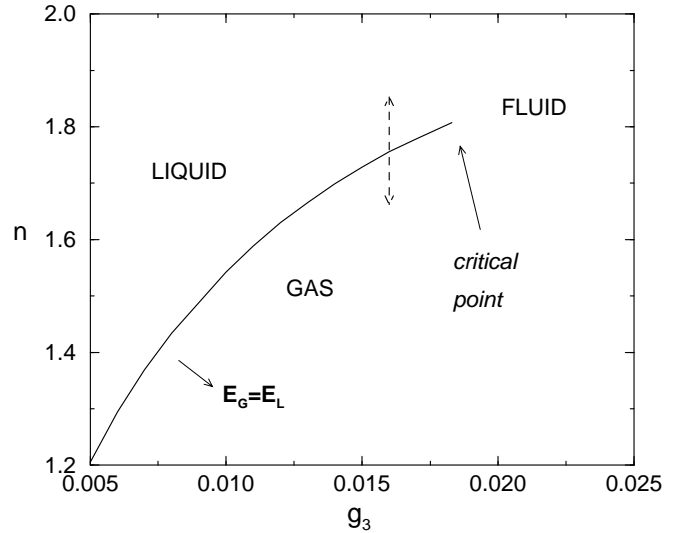}} 
\caption[dummy0] {Graphical representation of the interface of the
two distinct phase (gas and liquid), in the plane defined by the 
reduced number of atoms $n$ and the parameter $g_3$ (lower frame);
and for the central density $\rho_c$ versus $g_3$ (upper frame). 
The arrows in the lower frame correspond to the point where it occurs the phase
transition for $g_3=$0.016, when changing $n$.}
\end{figure}

For each $g_{3},$ the transition point given by the crossing point in the $E$
versus $n$ (see Fig. 2) corresponds to a Maxwell construction in the diagram of
$\mu $ versus $n$. At this point an equilibrated condensate should undergo a
phase transition from the branch extending to small $n$ to the branch extending
to large $n.$ The system should never explore the back bending part of the
diagram because, as seen in Fig. 2, it is an unstable extremum of the
energy. From Figs. 1-5, it is clear that the first branch is associated with
small densities, large radii, and positive chemical potentials while the second
branch presents a more compact configuration with a smaller radius a larger
density and a negative chemical potential.
This justify the term gas (G) for the first one and liquid (L) for the second one.
However we want to stress that both solutions are quantum fluids.

\section{Collective Excitations}
In this section, from the time evolution of GPG equation, 
given in Eq.~(\ref{NLS}), we consider the ground-state collective excitations 
for the system~\cite{stringa,SR,Clark}.
Following Ref.~\cite{Clark}, the collective excitations are described by 
the Bogoliubov equations~\cite{bogo,Pita,fetter,FR}. 
After including three body interactions they take the form
\begin{eqnarray}
\lbrack {\cal L_\nu}-\hbar \omega _{\nu }]u_{\nu }+
\{NU_{0}+2\lambda_3N^2|\psi_g|^2
\}[\psi_{g}]^{2} v_{\nu} &=&0 
\nonumber \\
\lbrack {\cal L_\nu}+\hbar \omega _{\nu}]v_{\nu}+
\{NU_{0}+2\lambda_3N^2|\psi_g|^2 \}
[\psi_{g}^{\ast}]^{2} u_{\nu} &=&0,
\label{bogo}\end{eqnarray}
where
\begin{equation}{\cal L_\nu}\equiv H_{0}-\mu+2U_{0}N\left|\psi_{g}
\right|^2+3\lambda_3N^2|\psi_g|^4 .
\end{equation} 
$H_0$ is the harmonic oscillator hamiltonian,
$U_0\equiv -({4\pi\hbar^{2}|a|})/{m}$, 
$\omega _{\nu}$ is the frequency of the collective oscillations,
$N$ is the number of atoms and 
$\psi_{g}\equiv \psi_{g}({\bf r})$ is the ground state solution of 
the Eq.~(\ref{SNLS}). 
The above equations have been solved by using several 
methods~\cite{SR,Clark,lewenstein}.
In the present calculations we have employed two methods:  
a time-dependent and a time-independent one. In the time-dependent procedure we
have added a weak perturbation to the potential and, with CN algorithm, 
examined the time evolution of Eq.~(\ref{NLS}) for a selected
point of the wave-function. 
The lowest collective oscillations ($\omega_\nu$) were determined through
Fourier transformation~\cite{Clark}.
By using the time independent algorithm, we have solved Eqs.~(\ref{bogo})
with the matching algorithm ~\cite{gior} generalized for two functions $u$ and
$v$. The method works by departing from the analytically known $u, v$ and
$\omega_\nu$ for the harmonic oscillator (chemical potential near to
3/2$\hbar\omega$).
Then we  successively apply the matching method for the coupled $u$ and $v$,
gradually decreasing the chemical potential. This allows to reach subsequent 
solutions, by employing the deformation algorithm described in Ref.~\cite{num}.  
We obtain exact agreement between both methods, time-dependent or 
time-independent one.

Figure 6 shows the collective frequencies $\omega_{\nu}$ as a function
of $n$ for the first mode ($l=0$). \ The solutions corresponding to
$g_{3}=0$ agree well with the ones given in Ref.~\cite{SR},
loosing stability as $\omega _{\nu}\rightarrow 0$. \ By using this criterium, 
we have obtained the regions of stability for $g_{3}=0.016$. For $g_{3}=0.03$
all the solutions are stable.
Following the inset of Fig. 6, for $g_3=0.016$, one can observe that,
as the number of atoms is increased, in the less denser phase, the
frequency of the collective excitations decreases and are related to
stable solutions till the point C$_{\rm G}$; from this point down
to the point A (increasing $n$), the frequency continues to decrease to
zero, but now related to meta-stable solutions.
As already explained previously in the discussion of Figs. 2-5, and also
from the variational energy solutions given in Fig. 1, it
is very likely that occurs a phase transition, from C$_{\rm G}$ to 
C$_{\rm L}$ 
(or from the meta-stable solutions, given in the branches C$_{\rm G}-$A and
B$-$C$_{\rm L}$, to the corresponding stable solutions with fixed $n$).
Once in the denser phase (from B passing through the point C$_{\rm L}$), the
frequency of the collective excitations increases as the number of atoms
increases, contrary to the behavior observed for the system in the less
denser phase.
This can be qualitatively understood considering the variational
energy of the two phases and the corresponding stable energy as shown in
Fig.~1. The curvature  of the variational energy as a function of
$\alpha$ at the minimum for the liquid phase is bigger than the
corresponding one in the gas phase [compare in Fig.1 the insets (I) and
(II) with the insets (III) and (IV)]. This indicates, in agreement
with Fig.~6, that the restoration force is stronger for the liquid phase
than for the gas phase and consequently the frequencies of the collective
modes starting at the point C$_{\rm L}$ are higher than the corresponding ones
for the gas phase ending at C$_{\rm G}$. As we include more particles the
frequencies of the oscillations increase in the liquid phase.
Corresponding to Fig. 6, in Fig. 7 the collective frequencies are shown
as a function of the chemical potential $\beta $. From right to left, as the
chemical potential decreases till C$_{\rm G}$, $\beta$ also decreases; from 
C$_{\rm G}$ to A, and from B to C$_{\rm L}$ the solutions are meta-stable, such that
the system will look for a transition to a stable branch (from C$_{\rm G}$, 
increasing $\beta$, and from C$_{\rm L}$, decreasing $\beta$). From C$_{\rm L}$, 
as we further decreases the value of $\beta$ the frequency of
the collective excitations increases.

\begin{figure}
\setlength{\epsfxsize}{1.0\hsize}
\centerline{\epsfbox{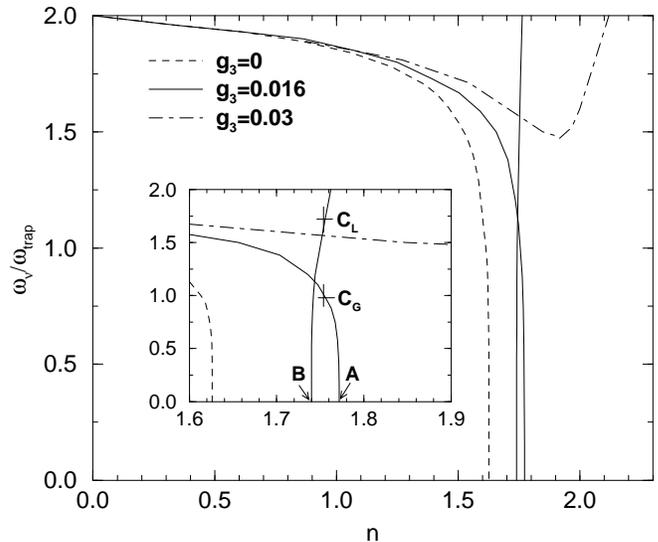}} 
\caption[dummy0] {Collective frequencies as a function of the reduced
number of atoms $n$. 
The inset shows the critical points  
corresponding to the previous figures.}
\end{figure}

\begin{figure}
\setlength{\epsfxsize}{1.0\hsize}
\centerline{\epsfbox{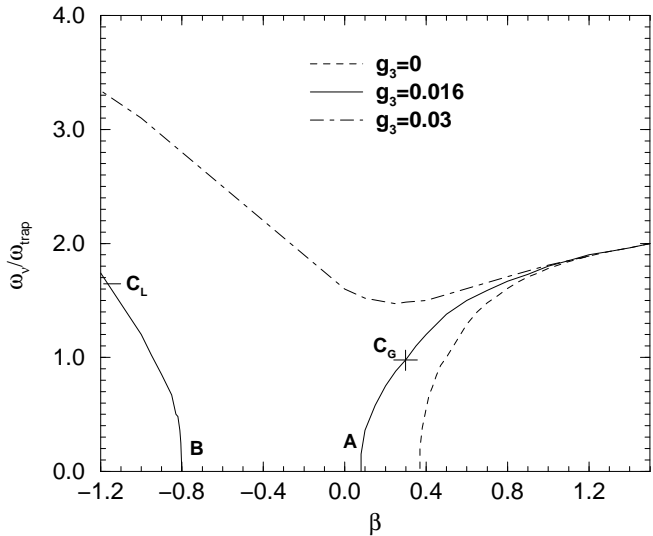}} 
\caption[dummy0] {Collective frequencies as a function of the chemical 
potential $\beta$.}
\end{figure}

\section{Conclusions}

To summarize, we have presented results for the total energy, chemical
potential, central density, in terms of the number of atoms in the
condensed state, for a range of values of the three-body strength. 
We also study the lowest collective mode excitations of the ground-state.
Our calculation presents, at the mean-field level, the consequences of a
repulsive three-body effective interaction for the Bose condensed
wave-function, together with an attractive two-body interaction. A
first-order liquid-gas phase-transition is observed for the condensed state
as soon as a small repulsive effective three-body force is introduced. In
dimensionless units the critical point is obtained when $g_{3}\approx 0.0183$
and $n\approx 1.8$. The characterization of the two-phases through their
energies, chemical potentials, central densities and radius were also given
for several values of the three-body parameter $g_{3}$. The results
presented in this paper can be relevant to determine a possible clear
signature of the presence of repulsive three-body interactions in Bose
condensed atoms. It points to a new type of phase transition between two
Bose fluids. Because of the condensation of the atoms in a single
wave-function this transition may present very peculiar fluctuations and
correlations properties. As a consequence, it may fall into a different
universality class than the standard liquid-gas phase transition, which are
strongly affected by many-body correlations. This matter certainly
deserves further studies.

{\bf Acknowledgments}

This work was partially supported by Funda\c c\~ao de Amparo \`a Pesquisa do
Estado de S\~ao Paulo and Conselho Nacional de Desenvolvimento Cient\'\i fico
e Tecnol\'ogico.


\begin{references}

\bibitem{bec} S.N. Bose, Z. Phys. {\bf 26}, 178 (1924);
A. Einstein, Sitz. Preuss Acad. Wiss. {\bf 261} (1924); 3 (1925).

\bibitem{PW} A.S. Parkins and D.F. Walls, Phys. Rep. {\bf 303}, 1
(1998); A. Griffin, D.W. Snoke, and S. Stringari, {\it Bose-Einstein
Condensation} (Cambridge University Press, Cambridge, 1995).

\bibitem{london} F. London, Phys. Rev. {bf 54}, 947 (1938);
F. London, {\it Superfluids II} (John Wiley and Sons, New York, 1954).

\bibitem{bogo} N.N. Bogoliubov, J. Phys. (USSR) {\bf 11}, 23 (1947).

\bibitem{and95}  M.H. Anderson, J.R. Ensher, M.R. Matthews, C.E. Wieman,
E.A. Cornell, Science {\bf 269}, 198 (1995).

\bibitem{brad97}  
C.C. Bradley, C.A. Sackett, J.J. Tollet and R.G. Hulet,
Phys. Rev. Lett. {\bf 75}, 1687 (1995);
C.C. Bradley, C.A. Sackett and R.G. Hulet, Phys. Rev.
Lett. {\bf 78}, 985 (1997); C.C. Bradley, C.A. Sackett, J.J. Tollet and R.G.
Hulet, Phys. Rev. Lett. {\bf 79}, 1170 (1997).

\bibitem{mew95}  
K.B. Davis, M.-O. Mewes, M.R. Andrews, N.J. van Druten, D.S. Durfee, 
D.M. Kurn, W. Ketterle, Phys. Rev. Lett. {\bf 75}, 3969 (1995);
M.R. Andrews, M.-O. Mewes, N.J. van Druten, D.S. Durfee,
D.M. Kurn, W.Ketterle, Science {\bf 273}, 84 (1996);
M.-O. Mewes,M.R. Andrews, N.J. van Druten, D.M. Kurn, D.S.
Durfee, and W. Ketterle, Phys. Rev. Lett. {\bf 77}, 416 (1996).

\bibitem{hidr} D.G. Fried, T.C. Killian, L. Willmann, D. Landhuis, 
A.C. Moss, T.J. Greytak, and D. Kleppner, Phys. Rev. Lett. {\bf 81}, 3811
(1998).

\bibitem{cornell} E. Cornell, J. Res. Natl. Inst. Stand. Technol.
{bf 101}, 419 (1996).

\bibitem{huang}  K. Huang, {\it Statistical Mechanics}, 2nd. edition (John
Wiley and Sons, New York, 1987).

\bibitem{rup95}  M. Edwards and K. Burnett, Phys. Rev. A{\bf 51}, 1382
(1995); P.A. Ruprecht, M.J. Holland, K. Burnett, and M. Edwards, Phys. Rev. 
A{\bf 51}, 4704 (1995).

\bibitem{baym96}  G. Baym and C.J. Pethick, Phys. Rev. Lett., {\bf 76}, 6
(1996).

\bibitem{bag99} J. Weiner, V.S. Bagnato, S. Zilio, and P.S. Julienne, 
Rev. Mod. Phys., {\bf 71}, 1 (1999).

\bibitem{kagan96}  
Yu. Kagan, A.E. Muryshev, G.V. Shlyapnikov, and J.T.M. Walraven, Phys.
Rev. Lett. {\bf 76}, 2670 (1996).

\bibitem{ADV}  N. Akhmediev, M.P. Das and A.V. Vagov, 
Int. J. Mod. Phys. B{\bf 13}, 625 (1999).

\bibitem{GFT}  A. Gammal, T. Frederico, and L. Tomio, {\it Trapped
Bose-Einstein condensed gas with two and three-atom interactions}, in
proceedings of the ``International Workshop on Collective Excitations in
Fermi and Bose Systems'', ed. by C. Bertulani, L.F. Canto and M. Hussein 
(World Scientific, Singapore, 1999).

\bibitem{esry}  B.D. Esry, C.H. Greene, Y. Zhou, and C.D. Lin, J. Phys. B 
{\bf 29}, L51 (1996).

\bibitem{josse}  C. Josserand and S. Rica, Phys. Rev. Lett. {\bf 78}, 
1215 (1997).

\bibitem{kagan}  Yu. Kagan, B.V. Svistunov, and G.V. Shlyapnikov, JETP Lett. 
{\bf 42}, 209 (1985).

\bibitem{fedi}  P.O. Fedichev, M.W. Reynolds, and G.V. Shlyapnikov, Phys.
Rev. Lett. {\bf 77}, 2921 (1996).

\bibitem{burt}  E.A. Burt, R.W. Ghrist, C.J. Myatt, M.J. Holland, E.A.
Cornell, and C.E. Wieman, Phys. Rev. Lett. {\bf 79}, 337 (1997).

\bibitem{efimov}  V. Efimov, Phys. Lett. {\bf B 33}, 563 (1970); Comm. Nucl.
Part. Phys. {\bf 19}, 271 (1990).

\bibitem{3atom}  T. Frederico, L. Tomio, A. Delfino, and A.E.A. Amorim, 
Phys. Rev. A{\bf 60}, R9 (1999).

\bibitem{gin}  V.L. Ginzburg and L.P. Pitaevskii, Zh. Eksp. Teor. Fiz. 34,
1240 (1958) [Sov. Phys. JETP 7, 858 (1958)]; 
E.P. Gross, J. Math. Phys. 4, 195 (1963).

\bibitem{Pita} L.P. Pitaevskii, Zh. \'Eksp. Teor. Fiz. {\bf 40}, 646 (1961)
[Sov. Phys. JETP 13,451 (1961)]. 

\bibitem{halo}  A.E.A. Amorim, T. Frederico, and L. Tomio, Phys. Rev. C {\bf 
56}, R2378 (1997).

\bibitem{hs}  M. Houbiers and H.T.C. Stoof, Phys. Rev. A {\bf 54}, 5055
(1996).

\bibitem{fw}  A.L. Fetter and J.D. Walecka, {\it Quantum Theory of Many -
Particle Systems} (McGraw-Hill, New York, 1971).

\bibitem{fetter}  
A.L. Fetter, Phys. Rev. A{\bf 53}, 4245 (1996).

\bibitem{num} A. Gammal, T. Frederico, and L. Tomio, Phys. Rev. E{\bf 60},
2421 (1999).

\bibitem{ames} W.F. Ames, {\it Numerical Methods for Partial Differential
Equations}, 3rd. ed., Academic Press, New York, 1992, pp. 111-115.

\bibitem{stringa} S. Stringari, Phys. Rev. Lett. {\bf 77}, 2360 (1996).

\bibitem{SR}  K.G. Singh and D.S. Rokhsar, Phys. Rev. Lett {\bf 77},
1667 (1996).

\bibitem{Clark}  
P.A. Ruprecht, M. Edwards, K. Burnett, and C.W.Clark, 
Phys. Rev. A {\bf 54}, 4178 (1996);
M. Edwards, P.A. Ruprecht, K. Burnett, R.J. Dodd, and C.W. Clark 
Phys. Rev. Lett. {\bf 77}, 1671 (1996). 

\bibitem{FR} 
A.L. Fetter, Ann. Phys. (N.Y.) {\bf 70}, 67 (1972);
A.L. Fetter and D. Rokhsar, Phys. Rev. A{\bf 57}, 1191 (1998).

\bibitem{lewenstein} L. You, W. Hoston and M. Lewenstein, Phys. Rev. A
{\bf 55}, R1581 (1997).

\bibitem{gior} N.J. Giordano, {\it Computational Physics}, Prentice-Hall,
New Jersey, 1997, pp. 257-272.

\end{references}
\end{document}